\documentclass[prl,twocolumn,epsfig,rotate,showpacs]{revtex4}
\usepackage{epsfig}
\usepackage{amsmath}
\usepackage{graphicx}
\usepackage{dcolumn}
\usepackage{bm}

\input{tcilatex}
\begin{document}

\title{Normal heat conduction in lattice models with asymmetry harmonic
interparticle interactions. }
\author{Yi Zhong }
\author{Yong Zhang}
\author{Jiao Wang}
\author{Hong Zhao}
\email{zhaoh@xmu.edu.cn}
\affiliation{Department of Physics and Institute of Theoretical Physics and
Astrophysics,Xiamen University, Xiamen 361005, China. }
\date{\today }

\begin{abstract}
We study the heat conduct behavior of a lattice model with asymmetry
harmonic inter-particle interactions in this paper. Normal heat conductivity
independent of the system size is observed when the lattice chain is long
enough. Because only the harmonic interactions are involved, the result
confirms without ambiguous interpretation that the asymmetry plays the key
role in resulting in the normal heat conduct of one dimensional momentum
conserving lattices. Both equilibrium and non-equilibrium simulations are
performed to support the conclusion.
\end{abstract}

\maketitle

The heat transport properties of low-dimensional systems have evoked
intensive studies for decades [1-13], aiming at to verify whether the
Fourier's law of heat conduction 
\begin{equation}
J=-\kappa \nabla T  \label{Four}
\end{equation}%
valid in low-dimensional materials. Here $J$ is the heat current, $\nabla T$
is the temperature gradient along the sample, $\kappa $ is the thermal
conductivity. At present, for momentum-conserving 1D fluids and lattices, it
is generally believed that the thermal conductivity should diverge as $%
\kappa \sim L^{\alpha }$ with the increase of the system size $L$ for
momentum conserving low-dimensional systems[14-19]. Meanwhile, some
counterexamples with size-independent thermal conductivities have been also
found, such as the rotator model [20,21], a 1D lattice in effective magnetic
fields [22], the variant ding-a-ling model [23].

Recently we find that momentum-conserved lattice models with asymmetric
interparticle interactions [24] can also result the normal heat conduction.
This result is confirmed by investigating the time-dependent behavior of
current autocorrelation functions in one-dimensional lattice systems with
the asymmetric and symmetric interaction potentials [25]. The current
autocorrelation is defined as 
\begin{equation*}
C(t)=\langle J(t)J(0)\rangle ,
\end{equation*}%
where $J(t)$ represents the current fluctuation at time $t$ and Here $%
\langle \cdot \rangle $ denotes the equilibrium thermodynamic average.
Following the linear response theory [27], the thermal conductivity may be
calculated by the Green-Kubo formula 
\begin{equation}
\kappa =\lim_{\tau \rightarrow \infty }\lim_{N\rightarrow \infty }\frac{1}{\
2k_{B}T}\int_{0}^{\tau }C(t)dt,
\end{equation}%
once the correlation function $C(t)$ is obtained, where $\tau $ is the time
of evolution, $L$ is the linear dimension of the system along which the
current flows, $k_{B}$ is the Boltzmann constant, $T$ is the temperature of
the system. Differing from the direct nonequilibrium calculation based on
equition (1), the thermal conductivity here is calculated with current
fluctuations in the euilibrium system. After studying different types of
interaction potentials, it is found [25] that with proper degree of
asymmetry, the current autocorrelation may show rapid decay which led to the
convergence of the Green-Kubo formula.

It is well-known that the asymmetry interaction may induce the thermal
expansion while the symmetry one may not, and real materials usually show
thermal expansion effect[26]. Thus, our finding has particular importance
for real materials. It implies that low-dimensional materials may also have
the size-independent thermal conductivity in the thermal limit as the bulk
materials, and the Fourier's law of heat conduction is generally valid also
for low-dimensional materials. However, as mentioned above, at present it is
generally accepted that the heat current autocorrelation decays in power-law
and the thermal conductivity diverges with the system size in
one-dimensional momentum conserving systems. Meanwhile, the models we have
studied [25,26] have a combination potential of nonlinearity and
asymmetricity. Therefore, whether the convergent thermal conductivity is
resulted by the nonlinearity or the asymmtry feature needs to be clearified. 
\begin{figure}[tbp]
\begin{center}
\includegraphics[angle=0,width=0.5\textwidth]{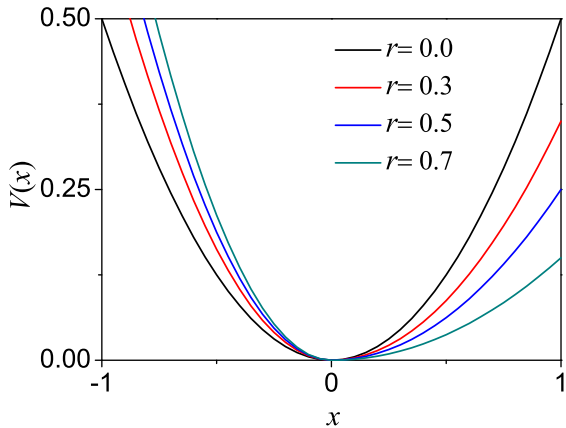}
\end{center}
\caption{plot of asymmetric harmonic interaction potential with $r=0.0$, $%
r=0.3$, $r=0.5$ and $r=0.7$. }
\end{figure}
\begin{figure*}[tbp]
\begin{center}
\includegraphics[angle=0,width=1.0\textwidth,height=5.5cm]{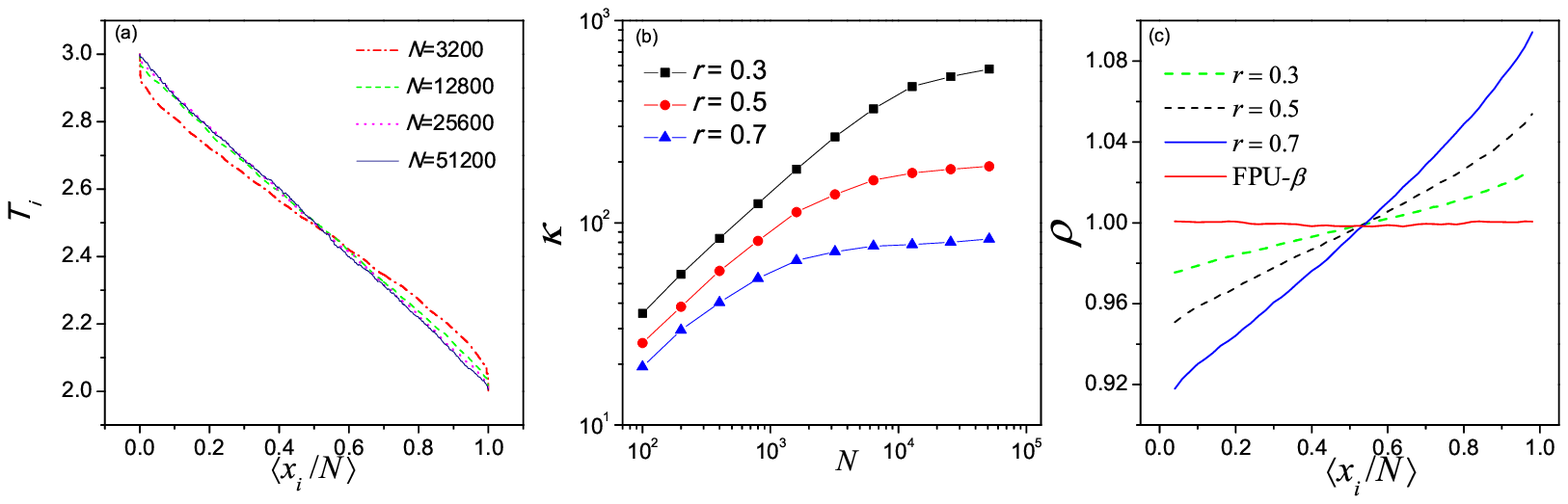} 
\end{center}
\caption{(a) Temperature profiles for the asymmetric harmonic interaction
potential with fixed $r=0.5$. The temperatures of the two heat baths coupled
to the system are $T_{L}=3$ and $T_{R}=2$ respectively, (b) The heat
conductivity $\protect\kappa $ vs the number of particles $N$ for $r=0.3$, $%
r=0.5$ and $r=0.7$, (c) The mass density function respectively for the
asymmetric harmonic interaction potential model with fixed $r=0.3$, $r=0.5$
and $r=0.7$ and the Fermi-Pasta-Ulam-$\protect\beta $ (FPU-$\protect\beta $)
model. The system size is $N=2000$. Other parameters are $T_{L}=3$ and $%
T_{R}=2$. }
\end{figure*}
This paper studies an one-dimensional momentum conserving lattice with
simple asymmetric interparticle interactions: The compress and stretch are
govern by different harmonic potentials. In more detail, we study the
lattice described by the Hamiltonian 
\begin{equation}
H=\sum_{i=1}^{N}{\frac{p_{i}^{2}}{2}}+V(x_{i}-x_{i-1}),
\end{equation}%
where $p_{i}$ and $x_{i}$ represent the momentum and the deviation from its
equilibrium position of the $i$th particle respectively. The potential is as
following: 
\begin{equation}
V(x)=\left\{ 
\begin{array}{ll}
{\frac{1}{2}}(1+r)x^{2} & \text{if $x<0$ ,} \\ 
{\frac{1}{2}}(1-r)x^{2} & \text{otherwise .}%
\end{array}%
\right. 
\end{equation}%
where $r$ control the degree of the asymmetry. The potentials with several $r
$ are plotted in Fig.1. This potential and its higher order derivatives are
continuous at $x=0$ except the second derivative.

To perform the nonequilibrium simulations, the Nose-Hoover heat baths [28]
with temperatures $T_{L}$ and $T_{R}$ are coupled to the left and right
particles respectively. The fixed boundary conditions are applied in the
simulation. The equations of the motion of particles in heat baths is given
by 
\begin{equation}
\dot{x}_{1,N}=\frac{p_{1,N}}{\mu },\quad \dot{p}_{1,N}=-\frac{\partial H}{%
\partial x_{1,N}}-\varsigma _{\pm }p_{1,N},\quad \dot{\varsigma}_{\pm }=%
\frac{p_{i}^{2}}{T_{\pm }}-1,
\end{equation}%
and the motions of $N-2$ other particles are described by 
\begin{equation}
\dot{x}_{i}=\frac{p_{i}}{\mu },\quad \dot{p}_{i}=-\frac{\partial H}{\partial
x_{i}}.
\end{equation}%
We integrate the equations of motion by using the leap-frog integrating
algorithm. The local temperature and local heat current at the $i$th site
are calculated by $T_{i}\equiv \langle p_{i}^{2}\rangle $ and $J_{i}\equiv
\langle \dot{x}_{i}\frac{\partial H}{\partial x_{i}}\rangle $
respectively[8]. The simulations are performed with a sufficient long time,
usually $t>10^{7}$, to ensure the system reaching a stationary state. In
such a state, the local current is equal to the global flux, $J_{i}=J$. To
avoid the finite-size effect, the system size $N$ is extended until the the
temperature profiles would fit with each other by rescaling the $x$ variable
with factor $1/N$, in which case $dT/dx\symbol{126}N/(T_{L}-T_{R})$ and the
thermal conductivity can thus be calculated following $\kappa =\langle
J\rangle N/(T_{L}-T_{R})$. 
\begin{figure}[tbp]
\begin{center}
\includegraphics[angle=0,width=0.5\textwidth]{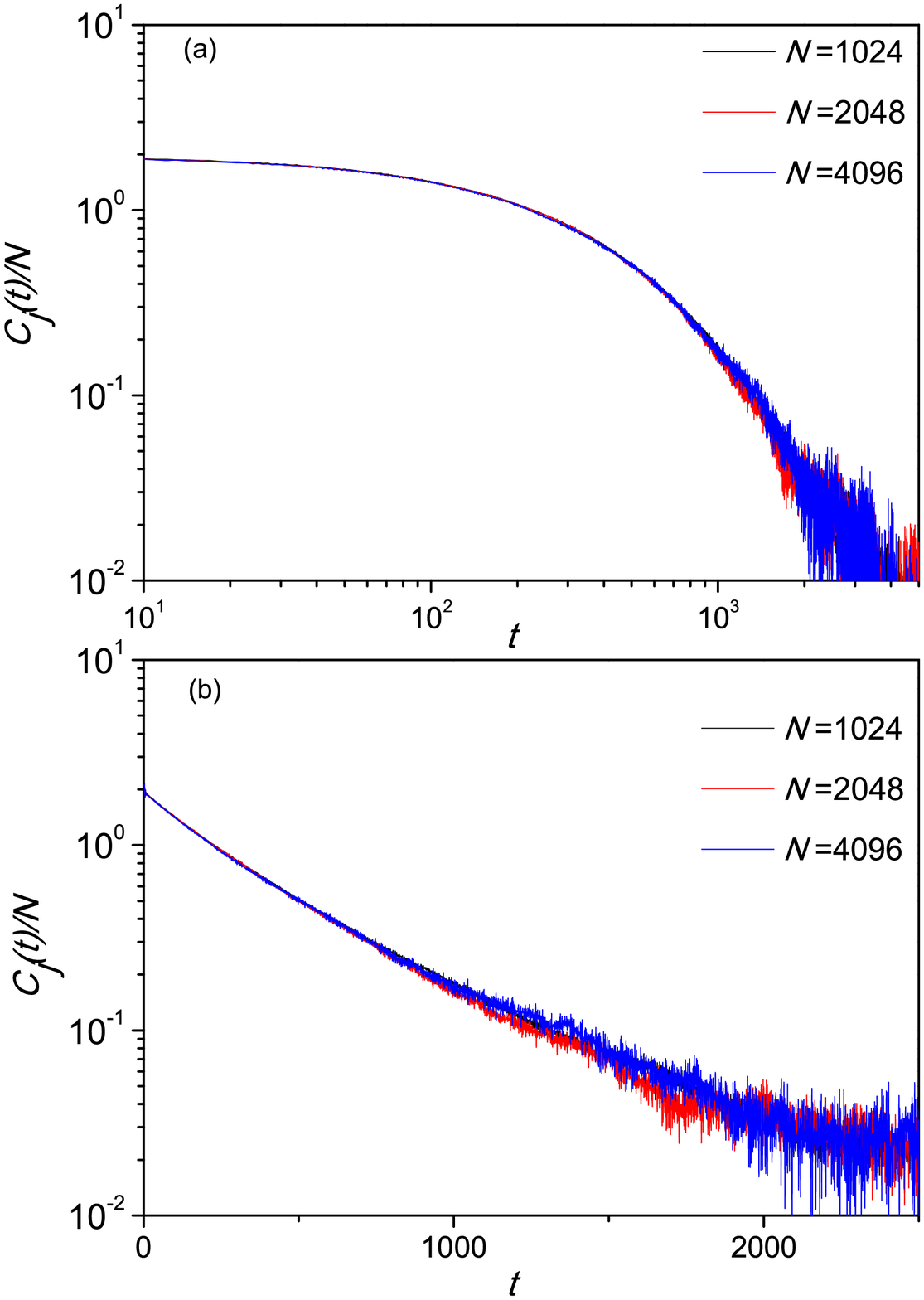}
\end{center}
\caption{The autocorrelation function of the heat flux,$C_{j}(t)$, for the
asymmetric harmonic interaction potential with $r=0.5$.}
\label{temp}
\end{figure}

Figure.2(a) shows the temperature profiles with $r=0.5$ for several system
sizes. One can see that for $N>10^{4}$ the temperature profiles are
well-rescaled together. Figure.2 (b) shows the thermal conductivity as a
function of the system size. It can be seen that $\kappa $ converges to be a
size-independent constant gradually. A remarkable difference to the case of
models with symmetry potential, such as the results shown in refs.
[4-7,10,24] where the finite-size effect disappears usually before $N>10^{3}$%
, is that the convergence threshold of the system size is quite long in this
model. This feature appears also in our previous work [24,25] where a
different asymmetry interaction potential is applied. We guss it may be one
of the reason why previous researchers have not observed convergent thermal
conductivity even they also investigated certain asymmetry-potential lattice
models.

The convergence threshold is related to the degree of the asymmetry. In
Fig.2, we also show the results with $r=0.3$ and $r=0.7$ respectively. It is
clear that the threshold of convergent thermal conductivity increases with
the decrease of the asymmetry degree. However, further increase $r$ may
result simulation difficult. We have to apply a very small intergral step to
guaratee the intergral precision.

As in reference [24], we calculate the mass density $\rho (x)$ along the
lattice chain at non-equilibrium stationary states. The results are shown in
Fig. 2(c). It can be seen that mass gradients do set up alone the chain in
the asymmetry cases, while in the sysmmetry case of $r=0$ there is no such a
gradient. The density is inverse proportional to the temperature. This is a
result of positive $r$, in which case the compress is difficult than the
stretch. If one apply a negative $r$ to the potential, in which case the
compress is easy than the stretch, he shall found that the mass desity is
proportional to the temperature. This is a qualitative different property
between symmetry and asymmertry lattice systems, and may provide clue to
understand why qualitative difference in heat transport is resulted. We guss
that the mass gradient can induce additional scattering of the current and,
together with other scattering mechanism, result the normal heat condduct
behavior. 

The decay behavior of the current autocorrelation can further confirm that
the asymmetry interparticle interactions may result the normal heat conduct
of lattice systems. To calculate the current autocorrelation function, we
first evolute the system for a sufficient long time to relax the system to
its equilibrium state. Then the currelation function $C(t)=\langle
J(t)J(0)\rangle $ is calculated by applying the emporal current fluctuations 
$J(t)=\sum\limits_{i=1}^{N}J_{i}(t)$. The decay behavior of $C(t)$
determines whether the heat flux violates the Fourier law of heat
conduction. If it decays as $C(t)\sim t^{-\gamma }$ with $\gamma <1$, the $%
\kappa $ will divergent following the Green-Kubo formula. If it decays
faster than $\gamma =1$, particularly with exponential decay of $C(t)\sim
e^{-\delta t}$, the Green-Kubo formula converges and the thermal
conductivity is size-indepedent in the termaldynamical limit. The Fourier
law is thus obeyed. Figure 3 shows the current autocorrelation functions
corresponding to the parameter sets applied in Fig. 2. To perform the
simulation, periodic boundary conditions are applied with several simulation
sizes. The temperature is set to be $T=2.5$ which is corresponding to the
average temperature applied in the nonequilibrium simulations. One may seen
that the curves with different simulation sizes overlap with each other,
indicating that the finite-size effect is avoided. It is clear, either with
the log-log plot or semi-log plot, the decay of the autocorrelation function
is quite fast, even approaches the exponential decay manner, indicating a
convergent thermal conductivity.

\bigskip In conclusion, the lattice model even with asymmetric harmonic
inter-particle interactions shows normal thermal conduction behavior. Our
nonequilibrium simulations obtain a size-independent thermal conductivity
when the simulation size is sufficient long. Our equilibrium simulations
show that the current autocorrelation decay faster than the power-law decay
of $C(t)\sim t^{-1}$, implying a convergent thermal conductivity according
to the Green-Kubo formula. Because this model involves only the asymmety
harmonic interactions, our results thus confirm that it is the asymmetry of
interaction potentials resulting the normal thermal conduct behavior of
one-dimensional momentum conserving lattices. 

This model has another obvious adventage. With a scale transformation $(%
\widetilde{x},\widetilde{t})\rightarrow (\alpha x,t)$, the Hamiltonian
changs as $\widetilde{H}\rightarrow \alpha ^{2}H$. Therefore, the dynamics
of the system keep unchange with the scale transformation. In more detail,
the systems with Hamiltonian $\widetilde{H}$ at temperature $\widetilde{T}$
and with Hamiltonian $H$ at temperature $T$ are identical, where $\widetilde{%
T}=\alpha ^{2}T$. Therefore, the conclusion of normal thermal conductivity
can be directly extended to any temperature. 

We have observed that the mass density is set up alone the lattice chain in
the case of asymmetry potentials. This phenomenon may be important in
understanding the microscopic mechanism of the normal thermal conductivity
in nonequilibrium systems. In equilibrium systems, there is no such a
stationary gradient of mass density. How a rapid decay of the current
autocorrelation is arisen is an open problem and asks further studies. 

We thank Shunda Chen for profitable discussions. This work is supported by
the NNSF (Grants No. 10925525, No. 10975115) and SRFDP (Grant No.
20100121110021) of China.

\end{document}